\documentclass[doublecol]{epl2} 

\usepackage{amssymb}
\usepackage{amsmath}
\usepackage{amsfonts}
\usepackage{amsopn}
\usepackage{graphics}
\usepackage{graphicx}
\usepackage{epsfig}
\usepackage{hyperref}
\usepackage{subfigure}

\title{Critical fluctuations of the electrical activity of the heart: Shortcomings of models of excitability and interpretation.}
\shorttitle{Critical fluctuations of fibrillating hearts.} 

\author{G. Attuel\inst{1} \and N. Derval\inst{2} \and T. Desplantez\inst{1} \and M. Haissaguerre\inst{2} \and M. Hocini\inst{2} \and P. Ja\"{i}s\inst{2} \and R. Dubois\inst{1}}
\shortauthor{G. Attuel \etal}

\institute{                    
  \inst{1} L'Institut de RYthmologie et mod\'{e}lisation Cardiaque. Universit\'{e} de Bordeaux, F33000 Bordeaux, France \\
  \inst{2} Centre Hospitalo-Universitaire du Haut L\'{e}v\^{e}que. CHU Haut L\'{e}v\^{e}que, F33600 Pessac, France
}
\pacs{87.19.Hh}{Cardiac dynamics}
\pacs{05.40.-a}{Fluctuation phenomena}
\pacs{05.70.Jk}{Critical point phenomena}

\date{\today}

\abstract{We report unexpected evidence of critical fluctuations of the electrical potential of the heart during atrial 
fibrillation in humans. Scale invariance and long range correlations are found, 
which we show cannot be accounted for solely with the property of excitability, since
disorder emerges by the formation of chaotic patterns in excitable media. 
To shed light on the data, we discuss the hypothesis that, in fact, fibrillation appears through a phase transition,
which we compare on phenomenological grounds to a quenched-in disorder magnetic transition. 
We infer that, during propagation of pulses, random pinning might occur
due to random modulation of the gap junction channels.}

\begin{document}
\maketitle

\section{Introduction}
Models of excitability describe the propagation of electrical pulses, called action potentials, 
which result from ionic exchange cycles between 
the cytoplasm of excitable cells and their extra-cellular medium.
Typical example are action potentials propagating through nerves \cite{Hodgkin52} 
and throughout the myocardium \cite{VdPol28} \cite{Noble62} 
\cite{Fitzhugh62} \cite{Nagumo62}. 
In their normal state, pulses can be generated that propagate by diffusion,
which takes place between cells through the gap junctions in the myocardium \cite{Thomas07}.
In its abnormal state, called arrhythmia, the myocardium is overwhelmed by 
rapid and irregular patterns of activation.
Atrial fibrillation (AF) is one important arrhythmia, illustrated in fig.(\ref{Trace}). 
The usual theoretical interpretation is based on self perpatuating spatiotemporal chaotic patterns in excitable media \cite{Winfree94}, 
as in the spiral wave break-up scenario \cite{Karma93} \cite{Fenton02}.
Fluctuations in those scenarios generically display Gaussian probability density distributions and short range correlations.
Experimentally, dynamical structures, so called "rotors", 
have been observed, and seem to pervade the tissue during fibrillation \cite{Jalife98}.
They are associated theoretically with spiral waves. 

In this letter, we give surprising experimental evidence of features of self organized criticality in AF in humans.
This is unexpected in cardiac dynamics, since it is supposed to lie in the class of spiral wave turbulence.
More specifically, we find non Gaussian probability density distributions of the fluctuating electrical activity, 
with heavy tails. Moreover, these fluctuations bear long range correlations.
In the classical framework of excitability, we show why large fluctuations are very unlikely. 
Thus, we postulate that AF is a state past a phase transition, 
of the kind found in magnetic systems with quenched disorder.
To illustrate this possibility, we outline a new interpretation, in which a major 
facet pertains to the cell to cell direct exchange of current through the gap junctions. 
We draw an analogy with pinned interfaces. Basically, cardiac pulses can be pinned
where cycles of neighboring cells become out of phase with one another, 
i.e. when gap junction channels become closed instead of open. 

\section{Measurements}


\begin{figure}[h!]
 \begin{center}
    \includegraphics[width=\columnwidth]{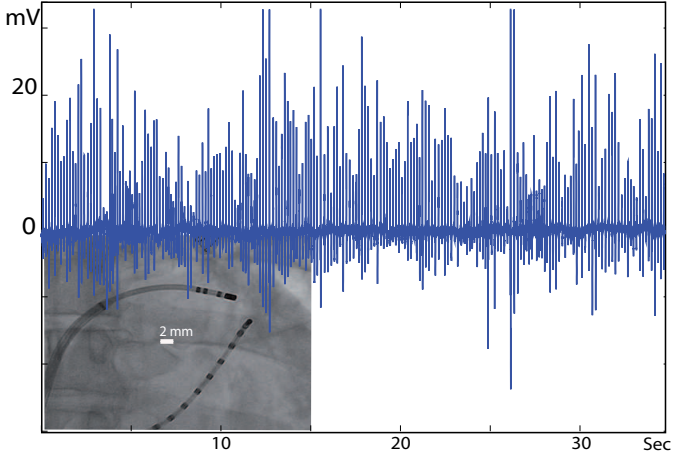}
 \end{center}
\caption{ 35 sec of paroxysmal AF are shown here.
The inset shows a radiography of the left atrium, where the bipolar electrode is located near the superior left pulmonary vein.
Another catheter comprising 10 electrodes runs along the coronary sinus.}
   \label{Trace} 
\end{figure}

\begin{figure}[h!]
 \begin{center}
    \includegraphics[width=\columnwidth]{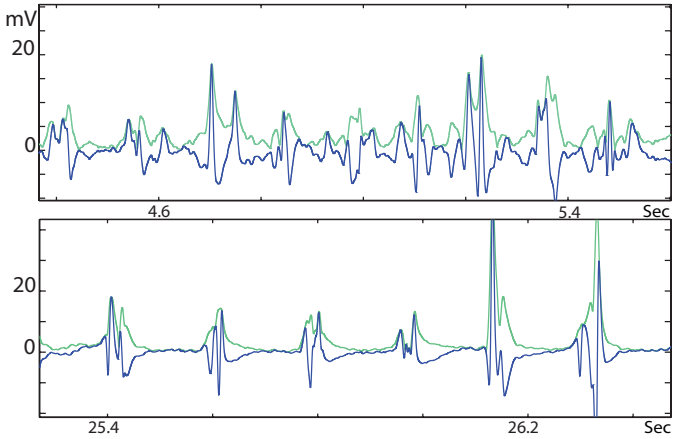}
 \end{center}
\caption{A close-up of a time series, where quiet baseline and fractionated signals during AF are shown, measured in the same patient. Their amplitude envelope $|A|$ are estimated using a Hilbert transform (in green).}
   \label{Frac} 
\end{figure}

Electrograms (egms), fig.(\ref{Trace}), are measured during AF, using 2 mm bipolar electrodes
in contact with the tissue, at a fixed location inside one of the human atria, held with a catheter by a medical practitioner.
The sampling rate is $1 kHz$. For practical reasons, towards 1 minute time series are considered,
consisting of about 5000 beats each. 
Two typical cases are drawn in fig.(\ref{Frac}), usually thought
to be very different in nature physiologically.
One is said to be fractionated, since at baseline
the voltage wanders a lot, while in the other case it exhibits isoelectric intervals.
Notably, we find no significant statistical differences 
between the two.

\begin{figure}[htbp]
\begin{center}
   \includegraphics[width=\columnwidth]{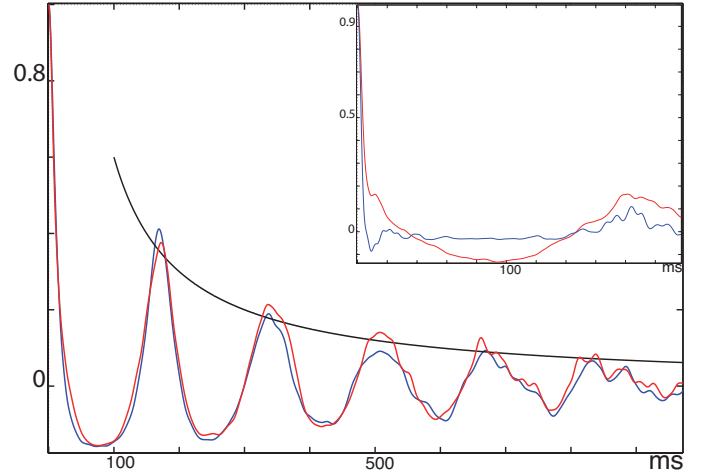} 
\end{center} 
\caption{Auto-correlation function of the amplitude of egms $|A|$.
An indicative solid line $\sim t^{-1}$ is drawn. In the inset, the auto-correlation of $A$ is plotted, where exponential decrease
is seen in the first $15 ms$.}
   \label{AutoCorrel} 
\end{figure}

As can be seen in fig.(\ref{AutoCorrel}), peaks appear naturally at about the mean cycle duration of the arrhythmia, 
typically ranging from $100 ms$ to $300 ms$ in humans, as opposed to about $1000 ms$ during normal sinus rhythm. 
Although the amplitude of each pulse seems to oscillate randomly, 
as confirmed by the exponentially decreasing auto-correlation function, the one of their envelope is decreasing asymptotically as a power law. 
The appearance of two well separated time scales suggests the formation of domains of coherent activity \cite{Goldenfeld92}.
In reaction-diffusion systems, algebraic decrease typically may be expected in 
the phase turbulence regime \cite{KPZ86} \cite{Chate96}, 
so that this finding alone is not inconsistent with the interpretation of meandering spiral waves \cite{Kalifa08}. 

\begin{figure}[htbp]
\begin{center}
   \includegraphics[width=\columnwidth]{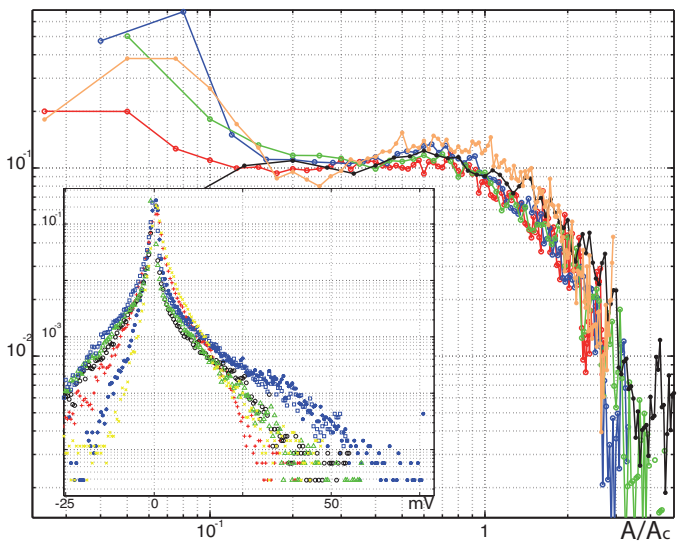} 
\end{center}
   \caption{In loglinear plot, the inset caption shows 
   a superposition of the normalized probability density distributions of egms A, from all over the left atrium of patient 2. 
   The high skewness is a hint of underlying mechanisms. They superpose in the form of eq.(\ref{eq:scaling_law}) as shown on the main graph in loglog plot.
    We choose to raise the number of boxes to 200, in order to visualize the statistical error.}
   \label{PDF} 
\end{figure}

\begin{figure}[htbp]
  \centering
  \hfill
   \subfigure{\includegraphics[width=\columnwidth, height=4cm]{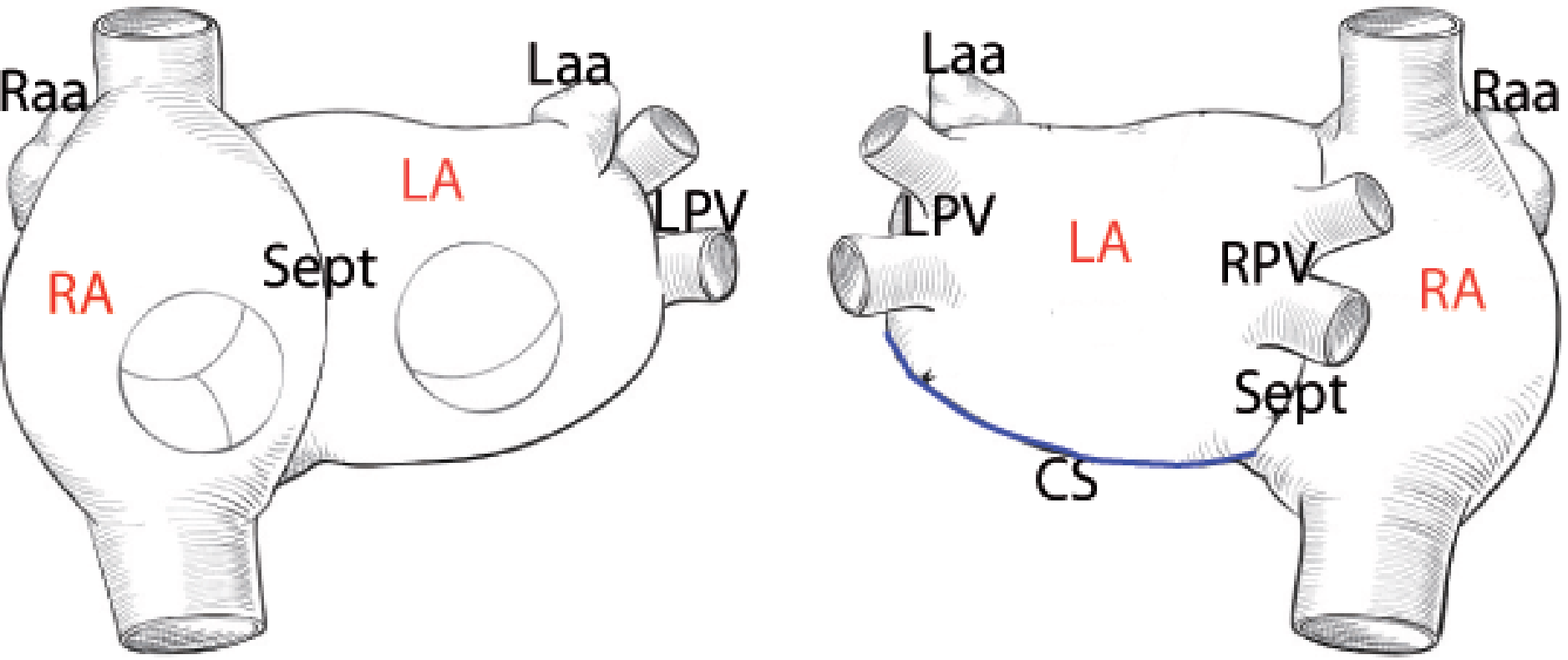}}
  \hfill
  \begin{tabular}{l|c|c|c|c|c|c|c}
  \hline
  N & RA & Raa & Sept & CS & RPV & LPV & Laa  \\
  \hline
  1 & & 1.4 & 1.5 & 1.8-2.4 & & & \\
  2 & & & 2.8 & 2.9 & 2.3 & 1.7  & 2.2  \\
  3 & & & & & &1.9 & \\
  4 &2.3 & & &2.2& 2.2 & 2.2 & \\
  5 &  & & 0.8  & 1.3 & 1.1 & 1.9 & \\  
  \hline
\end{tabular}
 \caption{Example of different stable values of $\tau$ found in the 5 patients reported here, where the recordings were done reliably. $\delta \tau \approx \pm 1$ except when otherwise mentioned. These values have been checked stable for about $10 mn$. Values $0.8$ and $1.1$ are flawed by a strong coherent component. Geometry of the atria. RA, LA: Left, Right atrium; LPV, RPV: Left, Right Pulmonary Veins. CS: Coronary Sinus. Laa, Raa: Left, Right appendage. Sept: Septum.}
 \label{Tau}      
\end{figure}

Nonetheless, fluctuations are large and their probability density distributions 
collapse altogether, showing scale invariance, as is seen
in fig.(\ref{PDF}). They can be cast into the form
\begin{equation}\label{eq:scaling_law}
P\left(A,A_{c}\right)=A^{-\tau}G\left(\frac{A}{A_{c}}\right)
\end{equation}
where $A$ is the egm varying amplitude in $mV$, $A_c$ is a cutoff and $\tau$ is a scaling exponent \footnote{We illustrate here when $A>0$, but it is also true for $A<0$ with slightly different exponents and cutoffs, and for $|A|$  as well with similar exponents.}. 
A study including 5 patients has been conducted, where various values of $\tau$ have been found. They range roughly between $1.3<\tau<3$ among patients, 
and span in general more restricted intervals in each patient taken individually. $\tau$ is found stable in time, as well as regionally, as summarized in table \ref{Tau}. $A_c = 40 \pm 20$ in the normalized units, for patients 1-3 and 5, and remains consistent within each patient. Patient 4 exhibited no cutoff, and patient 5 had a very regular right atrium, resulting in a pronounced deformation of the probability density distributions from right to left. Cutoffs are due to finite size effects, and the distance to the critical point. 
Since all atria are about the same size, it would be interesting to precisely ascertain 
the differences in the medical history, and characteristics of the heart, of patient 4 in that respect. Those issues will be reported elsewhere more precisely.

To our knowledge, similar fluctuations were not found in excitable systems, 
but are rather ubiquitous in complex systems. To name a few, they are found in such universal classes as that 
of directed percolation \cite{Hinrichsen00}, random field Ising model \cite{Sethna96}, or self organized criticality \cite{Bak87}, 
and experimentally in hard superconductors \cite{Urbach95} \cite{Olson97} \cite{Altshuler04}, neural networks \cite{Arcangelis06} \cite{Beggs03}, 
or in intermittent regimes of strong turbulence \cite{Ciliberto88}.
%

\section{Fluctuations}

We now specify why large fluctuations cannot easily emerge from purely excitable systems.
Generally speaking, in continuous media, reaction-diffusion systems are described by the evolution equation
\begin{equation} 
\frac{\partial}{\partial_t} u = \mathcal{R}(u) + D\Delta u
\end{equation}
where $u$ is a two component vector describing the reactant. One is the diffusive compound, 
while the other plays the role of a recovery variable. The nonlinear operator $\mathcal{R}$ 
describes the reaction to a perturbation of $u$, while $D$ is the diffusion coefficient. 
As concerns the rudiments of cardiac cell membrane dynamics, the membrane gating processes are
averaged out adiabatically \cite{VdPol28} \cite{Noble62}  \cite{Fitzhugh62} \cite{Nagumo62}. A mean-field 
phenomenological description \textit{\`{a} la} Landau thereof is interesting. 
A real scalar free energy, for the membrane potential $u_m$, takes the equivalent form
\begin{equation}\label{eq:Free}
\mathcal{F}_0\ =\  \int dx^{d} \left\lbrace - \alpha u_m^{2} + \beta u_m^{4} + \frac{1}{2}D \left(\nabla u_m\right)^{2} - \left(I -J_m\right)u_m \right\rbrace 
\end{equation}
with dissipative dynamics, $\frac{\partial}{\partial t} u_m = -\frac{\delta}{\delta u_m} \mathcal{F}_0$. 
$d$ is the dimension, $J_m$ represents the membrane current, positive when outgoing, the dynamics of which reads
\begin{equation} \label{J_m}
\frac{\partial}{\partial t} J_m = M \left(u_m\right) - b J_m + b J_0
\end{equation}
$J_0$ is a pacemaker current and $I$ is an external impulse of current. 
The function $M$, and the parameters $b$, $\alpha$, $\beta$ 
define $\mathcal{R}$, as the adiabatic response of the intrinsic more rapid time scales.  

On the one hand, periodic forcing, such as $I=I_0\cos\left(\omega t\right)$,
or some delay in the response $M\left(u_m\right)$ 
is sufficient for the appearance of phase locking, 
quasi-periodicity and chaos \cite{Parlitz87}  \cite{Glass97} \cite{Panfilov93}.
In fact, quasi periodicity has been observed, 
experimentally in animals and humans, associated with rotating patterns, 
similar to wandering spiral waves, at the onset of AF \cite{Jalife98} \cite{Garfinkel99}. 
On the other hand, scale invariance strongly suggests \textit{criticality}. 
Incidentally, phase transitions are possible in reaction-diffusion systems as shows eq.(\ref{eq:Free}), by varying 
one or more of the cell dynamics parameters, as observed for instance 
between different turbulent states of a modified Fitzhugh-Nagumo model\cite{Bar95}, 
very similarly to what happens in liquid-gas transitions.
To enter a more detailed analysis, let us suppose $M=a \mathbb{I}$, $a>0$ without any loss of generality.  
We shall consider the effect of $J_m$ which 
acts as an external field on $u_m$, but has its own dynamics.
Intersection of null-clines in $\mathcal{R}$ determines fixed points. 
The quiescent state $\frac{\partial}{\partial t}=0$, 
for $J_0=0$, $u_m= \pm \sqrt{\frac{\alpha - \frac{a}{b} }{\beta}}$, corresponds to $\alpha>\frac{a}{b}$. 
Whether $J_m$ is slaved to $u_m$ for $b>a$ is not essential here.
Transition from quiescent toward excitable states exists for a certain leaking current $J_0>J_{c0}$.
We limit ourselves to the case $\alpha \le \frac{a}{b}$, since we don't observe cardiac arrest (quiescence).
A transition from excitability to automaticity is achieved when $J_0 \geq J_{c1}$. 
In the excitable state, a finite perturbation of $J_m$ crosses the unstable manifold and initiates 
a homoclinic orbit, that visits the "up-state" and returns to the "down-state". 
A pulse results which propagates throughout the excitable medium via diffusion. 
In the automatic state, there are no stable fixed points, and pulses are generated regularly.
The conventional 
critical point $\alpha = 0$ is not physiological. One possibility
to have critical fluctuations is to set $\alpha=\frac{a}{b}$, where then $J_{c1}=0$. In that case indeed, on average
$\left\langle J_m \right\rangle= \frac{a}{b} u_m$, and an effective 
free energy may be considered, with $\alpha\rightarrow \alpha - \frac{a}{b}$, where then $\chi\sim\left(\frac{a}{b}-\alpha\right)^{-1}=\infty$.
The mean field response $\delta u_m$ to a perturbation in the current there reads
$ \delta u_m  =\chi \ \left(I-\delta J_m\right) $, where $\delta J_m$ is a deviation from the mean. When
$\chi\gg 1$, excitation combined with chaotic fluctuations $\left\langle I^{2} + \delta J_m^{2} \right\rangle$ 
can possibly magnify fluctuations in $u_m$ to become large. To confirm this analysis, however, measuring $J_m$
would be needed but is in practice difficult to perform.

In any event, the susceptibility is significantly large in
the Ginzburg region of parameters. It is here so narrow that it is unobservable in practice.  
In two dimensions using mean field critical exponents, we have
\begin{equation} 
\delta \ \epsilon \sim l_c^{2} D^{-1} \ll 10^{-12} 
\end{equation}
where $\delta \ \epsilon \equiv \delta \left( \frac{a}{b} - \alpha\right)$ 
measures the width of the critical region,
while $l_c \approx 1 nm$ is a cutoff length in the model, which is arguably 
equal to the Debye length of the cytoplasmic electrolyte, or perhaps the gap junction membrane thickness, only ten times as great, 
whereas the diffusive length, for a maximum intrinsic cell frequency of about 
$10Hz$, is \footnote{$D=\frac{\sigma}{\eta C_m} \approx 10^{-3} m^{2}s^{-1}$ is the potential diffusion, where $\sigma$ is the membrane 
linear conductance, $C_m$ its capacitance, and $\eta$ is the typical width of a cell membrane.} $l_D\sim D^{\frac{1}{2}} \approx 1 cm$ \cite{Plonsey07}.
This result coincides to a certain extent with the idea that dynamical phenomena
taking place at the gap junction level are of very high frequency, 
since a typical time scale for channel kinetics is $\lesssim 10 ms$,
which is averaged out as concerns pulse propagation. As a result, the only possibility is that the many degrees of freedom of 
spatiotemporal chaos in such systems have no long range correlations,
and that fluctuations are Gaussian, as one can notice
for instance in \cite{Bar95}.

As we have seen, all the types of fibrillation considered here contradict this fact.
We wish to review briefly a model which can be seen as adding interesting properties to models of excitability. 
This will establish a set of arguments leading to our postulate
singling out the role of the gap junctions in AF, as the work in \cite{Bub05} already suggests.

The homoclinic orbit makes it appropriate to approximate each cycle as $u_m=\mathcal{R}eal \left(|u_m|\exp(i\theta)\right)$.
$J_m$ remains positive and acts in the excitable case as a pinning potential for $u_m$, 
with mean value $J_m\approx\frac{a}{2b}|u_m|$. 
Also valid for any function $M$, after reducing to the phase variable $\theta$, one is
left with a potential energy of the form $H\cos\left(\theta\right)$ in the free energy,
therefore with the following phase model 
\begin{equation}
 \partial_{t}\theta =  D \Delta \theta - H\sin\left(\theta \right) + F 
\end{equation}
called the perturbed and over-damped sine-Gordon equation. It possesses all the required properties of excitability,
where $F$ defines a typical frequency derived
from $\mathcal{R}$, and $H= J_m |u_m|$, when $J_0=0$. 
A $2\pi$-kink soliton represents an action potential, with depolarizing and repolarizing fronts.

Now, if for any reason, yet to be found, the fluctuating part of $J_m$ takes over 
and falls randomly out of phase with $u_m$, with a lag that reads $\delta J_m \sim \exp\left(-i\phi\right)$, 
this amounts to the addition of impurities. The model becomes the Fukuyama Lee Rice model that
governs the behavior of charged density waves in impure magnetic 
materials \cite{Fukuyama78} \cite{LeeRice79} \cite{Rice76} \cite{Gruner88}. The equation then reads
\begin{equation}\label{eq:CDW}
\partial_{t}\theta = D \Delta \theta - h\left(\textbf{x}\right)\sin\left(\theta - \phi\left(\textbf{x}\right) \right) + F 
\end{equation}
where $h$ includes the fluctuations of the membrane current $\delta J_m$. 
The spatial distribution of $h\left(\textbf{x}\right)$ and $\phi\left(\textbf{x}\right)$ represent the random 
pinning strength and phase lag. There, strain  $D \Delta \theta +F$
yields to starting a cycle where ever it is greater than $\min \left\lbrace h\right\rbrace$ \cite{LeeRice79}. 
Strain then accumulates
in the immediate neighborhood, enhancing a cascade phenomenon. 
Eventually, salves of synchronous pulses of different sizes (domains) slide off, in the fashion of avalanches. 
For instance, the charge density wave velocity jumps, 
scaling like $\delta v \propto F^{\zeta}$ with $\zeta\neq 1$  \cite{Coppersmith90}.
Self organized criticality is typically found 
in those systems \cite{Fisher83} \cite{Myers93}, especially in the strong pinning case
$\frac{h}{D}>1$, with domain size distribution typically like eq.(\ref{eq:scaling_law}). 
The exponent $\tau$ is related to $\zeta$ \cite{Sethna96}. 

Any experimental measurement of the potential, associated with length scale 
$l$, is a linear superposition of the local order parameter and takes on the form $\tilde{A}\propto\frac{1}{N_l(x,t)}\sum_i^{N_l(x,t)}|u_m|\cos\left(\theta_i\right)=\mathcal{R}_{eal}\left(|\tilde{A}|\exp\left(i\varphi\right)\right)$,
where $N_l(x,t)$ is the number of the surrounding synchronous cells, 
since no contribution comes from the destructive interference which appears in the sum above $N_l$.
Summation over a domain of coherent activity does give indeed an amplitude proportional to the size of the domain.
$\tilde{A}$ will bear similar properties, as long as $l$ is much shorter than the cutoff length. 
Gaussian statistics will appear for events of smallest amplitude only, not exceeding scale  \footnote{
For instance $l\sim 1mm$, since a typical cell length is $100 \mu m$, and cell width is $10 \mu m$, 
with $z$ is the average number of connected cells,
$z\approx 10$ in normal tissue, therefore $N_l > 1000$} $l$.

\section{Gap junctions}
\begin{figure}[htbp]
 \begin{center}
    \includegraphics[width=.8\columnwidth]{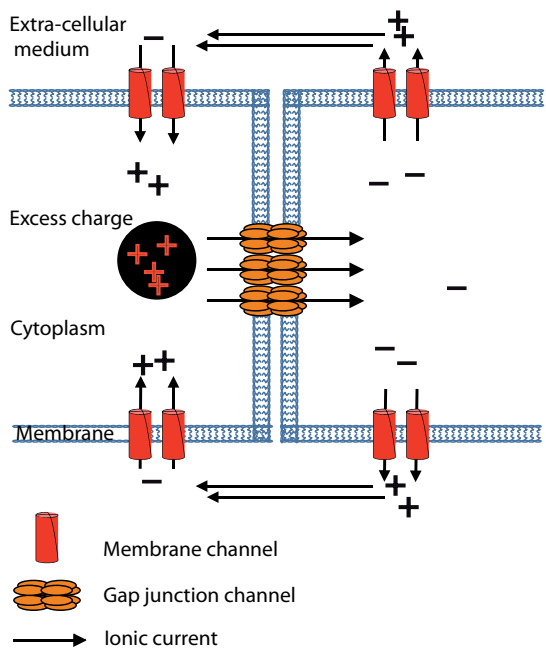}
 \end{center}
\caption{A sketch of the exchange of positive ions at the depolarizing front, through the gap junction channels, and across the membrane channels via the extra-cellular medium (arrows). The hypothesis that we formulate is of an instability of the coupling between local excess of positive charge and the kinetics of the gap junction channels that is capable of driving macroscopic excess, which in turn perturbs the kinetics of the membrane channels of the whole cell.}
   \label{gaps} 
\end{figure}

Since a chaotic attractor is the compact closure of all unstable periodic orbits \cite{Eckmann85} \cite{Auerbach87}, 
it seems very unlikely once again that such random
phase lag between $u_m$ and $J_m$ would emerge from chaos solely.
An easy way to bring on random phase lag is if 
random modulation of the opening and closing of the gap junction channels occurs.
Let us assume so, if the modulation becomes strong enough, at times, 
so as to create a local current source-sink mismatch through the gap junctions, 
there will be some random lag in the activation time of the next cell,
or in other words, a random modulated phase resetting of the cell cycles.
Under normal circumstances, positive charge diffuses through 
the gap junction and triggers the membrane channels of the next cell, 
facilitating the propagation of the front. Unknown is what happens if 
a macroscopic excess of positive charge sits before the gap junction channels, see fig.(\ref{gaps}). 
In a follow-up work, we further explore a possible instability of the local accumulation of ionic charges,
close to the gap junctions, that creates local incoherent modulations of their channel kinetics. 
An energy gap of chaotic collective modes appears
$h(x,t)\exp\left(-i\phi(x,t)\right)$, coupled to the dynamics of each cell. 

\section{Discussion}

Since they characterize here the local properties
of the electrical activity, the non universal values of $\tau$  might be related to the anisotropy of propagation,
just like in directed percolation \cite{Albert98}, or more generally to long time memory effects \cite{Grassberger97}.
The presumed physiology describing AF is a state 
of multiple "wavelets" of action potentials, which self sustain \cite{Allessie79}. It is long believed that
a critical number of such independent "wavelets" are necessary to sustain the electrical activity \cite{Moe59}. 
These ideas fuel the concept of so called fractionation, as being a genuine signature 
of the more numerous wavelets locally active near the probe.  
We have suggested on the contrary, that fractionated signals seem wrongly 
differentiated from more isoelectric baselines, since
both cases bear the same statistical properties. 
Besides, interestingly, crucial aspects of AF are left aside 
in this classical framework. One such crucial property is the remodeling of AF, which could be related to long time memory effects.  
The term refers to the worsening of AF in the sense that spontaneous conversion to 
sinus rhythm is less and less likely as time goes by \cite{Allessie95}. 
The phenomenology of pinning at the gap junctions is capable of rendering such a feature per se.
Remodeling, in that case, is the process of transiently redistributing 
the accumulated excess charge throughout the medium.



\section{Conclusion}
We have identified features of self organized criticality in human AF, which, we believe, rules out
the standard approach of AF, consisting of purely excitable reaction-diffusion processes.
In a broad sense, since statistical properties testify the universal class of underlying mechanisms,
we have given phenomenological arguments to the introduction of 
the gap junction channel dynamics as being essential, 
especially in addressing electrical remodeling. 

\acknowledgments
We are thankful to P. Attuel, O. Bernus, M.-C. Firpo, L. Glass, J. Kruithof, E. Vigmond, and M. Wyart for critically reviewing versions of the manuscript.  G. Attuel thanks E. Vigmond for interesting discussions, and is grateful to P. Attuel for introducing the field of AF to him.
The research leading to these results has received partial funding from the European Union Seventh Framework Programme (FP7/2007-2013) under Grant Agreement HEALTH-F2-2010-261057, partial financial support from the "Prix Coumel" of the Soci\'{e}t\'{e} Francaise de Cardiologie, and benefited from state aid managed by Agence National de la Recherche, in accordance with Recherche Investissements Avenir ANR- 10-IAHU-04.

\bibliographystyle{unsrt}
\bibliography{AF_CritiQ.bib}

\end{document}